\documentclass{article}
\usepackage{xcolor}
\usepackage{amsmath}

\usepackage[utf8]{inputenc}
\usepackage{graphicx}
\graphicspath{ {./images/} }
\usepackage{geometry}   
\usepackage{graphicx,epsfig}
\usepackage{amssymb}
\usepackage{float}
\usepackage{authblk}
\usepackage{indentfirst}
\usepackage[numbers,sort&compress]{natbib}
\usepackage[skip=2pt,font=scriptsize]{caption}
\begin{document}

\title{Tailoring high precision polynomial architected material constitutive responses via inverse design}
\author[1]{Brianna MacNider}
\author[1]{Ian Frankel}
\author[1]{Kai Qian}
\author[2]{Alan Pozos}
\author[3]{Aketzali Santos}
\author[4,5]{H. Alicia Kim}
\author[1,5]{Nicholas Boechler*}

\affil[1]{Department of Mechanical and Aerospace Engineering, University of California San Diego, 9500 Gilman Drive, La Jolla, CA 92093}
\affil[2]{Unidad Profesional Interdisciplinaria en Ingeniería y Tecnologías Avanzadas, Instituto Politécnico Nacional, Mexico City, Mexico.}
\affil[3]{Escuela Superior de Ingeniería Mecánica y Eléctrica Unidad Culhuacan, Instituto Politécnico Nacional, Mexico City, Mexico.}
\affil[4]{Department of Structural Engineering, University of California San Diego, 9500 Gilman Drive, San Diego, CA 92093, hak113@ucsd.edu}
\affil[5]{Program in Materials Science and Engineering, University of California San Diego, 9500 Gilman Drive, San Diego, CA 92093}
\date{}

\maketitle

\begin{abstract}

The design of specified nonlinear mechanical responses into a structure or material is a highly sought after capability, which would have a significant impact in areas such as wave tailoring in metamaterials, impact mitigation, soft robotics, and biomedicine. Here, we present a topology optimization approach to design structures for desired polynomial nonlinear behavior, wherein we formulate the problem in such a way as to decouple the nonlinear response from the stiffness. We show results across qualitatively different polynomial behaviors while achieving a high degree of precision, creating a path toward analytically tractable nonlinear dynamical systems. The approach enables access to previously difficult to design for, or hitherto unachieved, nonlinear behavior via optimized structures, which can furthermore be incorporated as unit cells of designer materials with tailored nonlinear properties.

\end{abstract}

\section{Introduction}

Nonlinear behavior has been increasingly exploited to achieve useful mechanical effects. Recent work in acoustic, or ``phononic'', metamaterials makes extensive use of unit-cell-scale structural features that exhibit nonlinear force-displacement laws to manipulate waves, inducing novel behavior such as dispersion tuning, harmonic generation, nonreciprocity, and solitary and rarefaction waves \cite{patil_review_2022}. Explorations into impact and vibration mitigation, energy trapping, and energy harvesting have used nonlinearity to achieve drastic performance improvements \cite{shan_multistable_2015,liu_recent_2015,harne_review_2013}. Applications which involve large deformation regimes, such as soft robotics and bio-interface systems, indeed rely upon the nonlinear response of their materials \cite{holmes_snapping_2007,james_layout_2016,ray_bio-integrated_2019}. Further, in many mechanical signal processing and logic strategies, nonlinear responses are central to their functionality \cite{boechler_bifurcation-based_2011,song_additively_2019,el_helou_digital_2021,yasuda_mechanical_2021}. However, exploration of the role of different types of mechanical nonlinearities in such systems has been limited, due, in large part, to the lack of a general design approach for physically realizing desired nonlinear constitutive properties.

One approach for achieving tailorable nonlinear mechanical properties is to use geometric nonlinearity (\textit{e.g.} large deformation or rotation, but small strain, effects). An important point to note is that, within the context of metamaterial-like structured materials, the use of geometric nonlinearity allows the mesostructural geometric nonlinearity to transfer to an effective ``material'' nonlinearity (typically resulting in more nonlinear tailorability than, \textit{e.g.}, chemical material synthesis methods). As such, the ability to tailor the nonlinear force-displacement response of structures can be considered applicable to both material and structural nonlinear response design problems. 

While discrete classes of nonlinearity have been studied in structured metamaterials (\textit{e.g.}, specific cases of cubic, quadratic, Hertzian, etc.), there has been little physically realizable exploration into the spaces in \textit{between} these classes (\textit{e.g.}, small variations to these nonlinear laws, or smooth variations between them), such that many potentially interesting nonlinear effects are currently restricted to analytical or computational treatment only. While several mechanisms have been explored for their ability to be (passively) tuned within their local nonlinear vicinity, including contact nonlinearities \cite{porter_granular_2015}, snap-through based structures \cite{cao_bistable_2021}, and tensegrity based structures \cite{fraternali_multiscale_2014}, such tunability is highly limited, being dependent upon specific mechanisms in each case which are not generally, widely applicable and which possess a finite range of nonlinearity. Indeed, there currently exist few methods to design a structure that will achieve a broadly-tailorable and \textit{highly precise}, desired mechanical nonlinear response. 

Several inverse design approaches to alter a structure’s force-displacement curve have been previously explored. Topology optimization has been used to tailor a small number of force-displacement control points, often in the pursuit of designing snap-through structures \cite{james_layout_2016,bruns_toward_2004,bhattacharyya_design_2019,bruns_numerical_2002}. Improved energy absorbing materials have been designed by optimizing to increase the area associated with hysteretic paths traced by varied force-displacement curves \cite{deng_topology_2020,chen_design_2018,https://doi.org/10.1002/advs.202204977}. A few works have explored the problem of optimizing a constitutive law over large strain ranges using trusses \cite{wang_design_2014}, splines \cite{jutte_design_2008,jutte_design_2010}, and high fidelity topology optimization \cite{CurveTracing2001,lee_enforcing_2020,li_design_2021,li_digital_2022}. Recent work has also investigated the use of machine learning to seek specific force-displacement curves, however it sometimes resulted in imprecise matches with the desired responses, and often requires large training data sets \cite{deng_inverse_2022, vlassis_denoising_2023,ha_rapid_2023,brown_deep_2023,zheng_deep_2023, Kochman_paper}. Our investigations found that seeking tailored nonlinear mechanical properties by directly targeting force or strain energy versus displacement curves (as used in Refs. \cite{wang_design_2014,jutte_design_2008,jutte_design_2010,CurveTracing2001,lee_enforcing_2020,li_design_2021,li_digital_2022,deng_inverse_2022, vlassis_denoising_2023}) unnecessarily restricts the design space and, hence, the achievable range of nonlinearities, by introducing an implicit coupling between stiffness and nonlinearity. This issue has previously been handled by carefully selecting target force-displacement curves which exist in compatible stiffness-nonlinear regimes (considering boundary conditions, minimum feature sizes, and constituent material) \cite{CurveTracing2001,lee_enforcing_2020,vlassis_denoising_2023}, and through the use of multiple materials to allow for a wider range of achievable overall structural stiffness (which adds significant, sometimes prohibitive, challenges for practical realization \cite{li_design_2021,li_digital_2022}). 

Here, we present a general approach for designing two-dimensional (2D), single material, nonlinear structures for a high precision one-dimensional (1D) nonlinear mechanical response. As schematically shown in Figure \ref{MethodOverview} A-C, we employ a level set topology optimization method, combined with mechanical intuition in the design of initial conditions, and target ratios of fitted polynomial coefficients as the objective function. Considering first the polynomial objective, we note that, particularly in the context of nonlinear dynamical systems, polynomial potentials \cite{patil_review_2022} are of particular interest due to their analytical tractability. Further, the exact type of polynomial response qualitatively affects the system response, for instance while both are ``stiffening'' in tension, a quadratic and cubic nonlinearity generate solely second versus third harmonics, respectively \cite{NonOscBook}. Similarly, works such as Ref. \cite{chaunsali_self-induced_2019} necessitate a precise balance between two nonlinear curves (in contrast to other, more forgiving, nonlinear objectives such as ``plateau-like'' behaviors sought after in low speed impact absorption). Next, considering the targeting of ratios, this approach allows us to decouple the optimization problem from the magnitude of structural stiffness within each design iteration, enabling the design to soften or stiffen as needed to achieve the desired degree of nonlinear behavior. Thereafter, the design can be scaled by adjusting the constituent material modulus, or size of the structure as is shown in Figure \ref{MethodOverview} C-E. This approach focuses the optimizer solely on the shape of the nonlinear response, circumventing the complexities of stiffness-nonlinearity dependencies. As such, it simplifies the inherently highly nonlinear design space to allow successful inverse design for targeted nonlinearities with precision. In other words, it is capable of exploring more design options and reliably finding solutions to a wider range of nonlinearities. Finally, the use of the level set method provides us with only the binary and smooth designs, enabling immediate manufacturing without post-processing, maintaining precision of the design (as opposed to density-based approaches, which often require post-processing and may lose some design precision in the process). The optimized nonlinear structures found herein can be incorporated into microstructures, lattices, or chains of springs and masses to form effective bulk material nonlinearities in pursuit of targeted designer materials. Geometric nonlinearity can be treated as an effective material nonlinearity through consideration of a system of multiple unit cells, consisting of repeated optimized structures, as contributing to an effective, homogenized, higher scale bulk behavior (see Figure \ref{MethodOverview}E for an example).  We demonstrate the successful optimization of structures for multiple classes of nonlinearities achieved in the prior literature without employing the same microstructural mechanisms typically associated with said nonlinearities. In addition, we show a semi-continuous sweep across the linear plus cubic design structural response space, demonstrating the ability to fine tune structures {reliably for highly specific forms of nonlinearity, rather than only broad classes. 

\begin{figure}
    \centering
    \includegraphics[width = \textwidth]{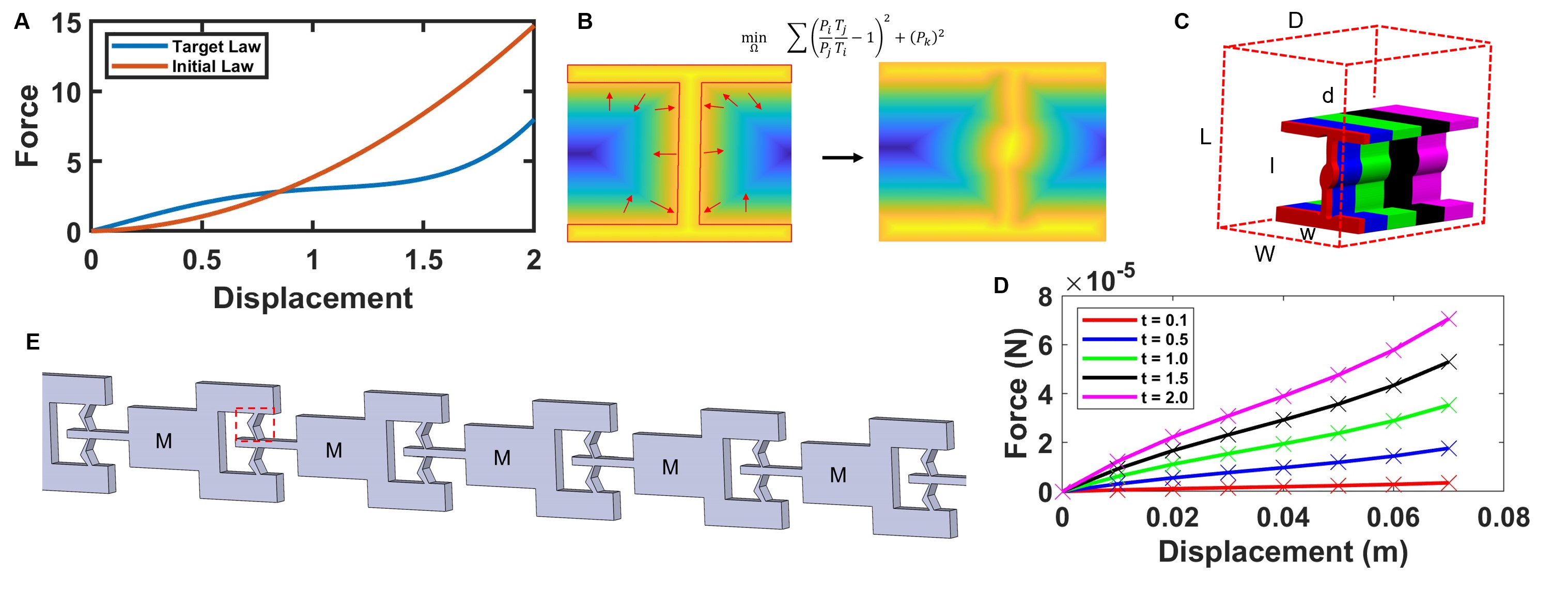}
   \caption{An overview of the proposed methodology used to design structures for targeted nonlinear mechanical behavior. A-B) A visualization of the level set approach for targeting coefficient ratios of a polynomial fitted to the force-displacement law shown in Panel A noting that the initial law shown is arbitrary, and depends upon the chosen initial condition, while the target law shown is the softening-to-stiffening target law displayed in Figure \ref{FigExamples}C). The solid red line in Panel B indicates the location of the zero level set, or the boundary of the structure, which is updated with each iteration as the design is changed. Blue and green indicate negative level set values (points outside of the structure), while yellow indicates positive level set values (points inside of the structure). Equation~\ref{Eq1}, the formulated objective, is shown in Panel B. C-D) A demonstration of how the stiffness of a structure (C) can be scaled by varying the out of plane thickness without affecting the shape of the nonlinearity (D), enabling designing for nonlinear behavior independent of stiffness, while still allowing for the modulation of global structural stiffness post design process. E) A visualization of how an optimized structure might be incorporated as a nonlinear spring in an example spring-mass chain (in which M refers to large masses, relative to the mass of the springs), forming an effective bulk nonlinear material. We note that this example is non-exhaustive, and that there are many ways in which such a nonlinear spring could be incorporated into a bulk nonlinear material.}
    \label{MethodOverview}
\end{figure}
    
\section{Results}
\subsection*{Heuristic Nonlinear Spring Design Method}

As a first step to our general design methodology, we examine how to achieve broad classes of nonlinearity in a structure. Qualitatively, a softening nonlinearity can be achieved by transitioning a beam-like structure via large deformation and rotation from an axial-dominated loading regime to a bending dominated loading regime, while a stiffening nonlinearity can be achieved by reversing this transition (see, \textit{e.g.}, Figure~\ref{Fig2} A). By tailoring the shape of the structure, we can adjust both the amount of softening and stiffening and where these regimes exist in the deformation range. This can be done in either a single ``spring'' (in which, herein, we refer to a beam-like structure undergoing large deformation as our spring), or through a combination of multiple springs in series or in parallel forming an effective spring. We note that within the context of linearized stiffnesses, the stiffness of the overall structure will be dominated by the stiffness of the softer spring for springs in series, while the stiffer spring will dominate for springs in parallel. The resultant behavior can thus be tailored to produce the desired type of nonlinear behavior in the desired displacement or force region.

\begin{figure}[]
    \centering
    \includegraphics[width = \textwidth]{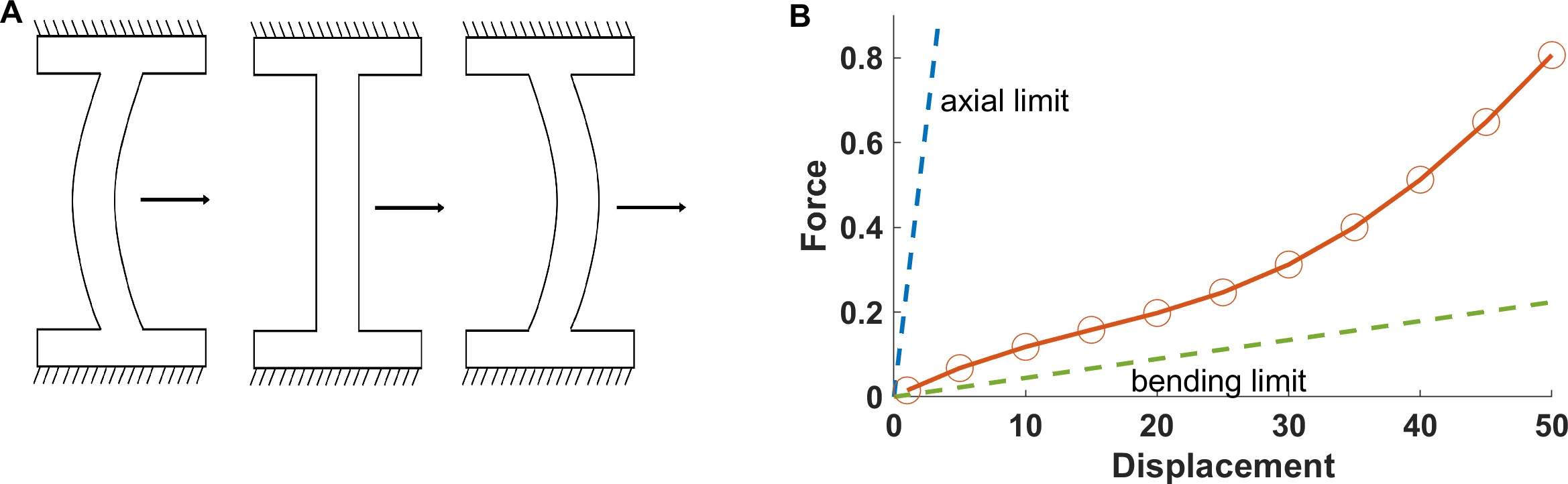}
   \caption{A demonstration of the mechanisms of geometric nonlinearity and their associated small deformation limits. A) A beam which uses geometric nonlinearity to achieve a softening to stiffening nonlinearity. B) The force-displacement curve of the beam in panel A (orange, circles), along with the limits calculated using the maximum and minimum stiffness from a full axial (top dashed line) and full bending (bottom dashed line) linear loading regime, respectively. These limits assume no negative stiffness in the structure.}
    \label{Fig2}
\end{figure}

Parameters of importance can be identified by examining the case of a straight or curved beam being pulled through a fixed displacement range as shown in Figure~\ref{Fig2} A. A spring of this general design could easily be incorporated into a spring-mass chain, with a mechanical frame present to enforce rigid boundary conditions, as displayed in Figure \ref{MethodOverview} E. Considering a system of this design, we take the beam highlighted in the dotted red box in Figure \ref{MethodOverview} E as our design domain. There exist inherent limits in how stiff or soft this structure can be made, and similarly, the magnitude of forces that can be reached in a given displacement range. Consider first the blue dashed upper line in Figure~\ref{Fig2} B, which corresponds to the limit $EAL^{-1}$ and describes linear stiffness corresponding to the axial deformation of a long, thin elastic beam, where $E$, $A$, and $L$ are the beam's elastic modulus, cross sectional area, and length, respectively. This limit can be interpreted to define the approximate maximum force that could be achieved at a given deformation $\delta$, by pulling a beam that extends normally from a fixed boundary, in an axial fashion. Such a regime could be imagined to be achieved when the beam in Figure~\ref{Fig2} A is pulled far to the right ($\delta >> L$). This further serves as a small ($\delta << t$, where $t$ is the beam width) and very large deformation ($\delta >> L$) limit on the structural stiffness, with implications on for the effect of volume ratio in an inverse design method on achievable stiffnesses. An exception to these limits, as concerns local stiffness at intermediate displacements is the case of snap through and snap back (or snap down), where infinitely positive, infinitely negative, and zero local stiffnesses can be achieved \cite{BazantBook}. For instance, considering a structure that exhibits a single snap through or snap back feature, we could expect the following. For both a snap-through and snap-back structure, there is guaranteed to be two points of zero stiffness. For a snap-through structure, there is also guaranteed to be a point of finite negative stiffness which approaches a vertical slope in the limit of snap-back behavior. This vertical slope could be considered as an infinitely positive or negative stiffness. For a snap back structure, there is similarly guaranteed to be two points of vertical slope. Similarly to the blue dashed line forming the upper limit, we can define an approximate lower bound, denoted by the green dashed line in Figure~\ref{Fig2} A for stiffness in small deformation ($\delta << t$) regimes, which corresponds to the $EIL^{-3}$ bending stiffness of a long thin beam, with $I$ being the second moment of area. The exception to this bound would be to include contact effects, or to include a pre-load on a structure exhibiting negative stiffness onset. Without considering contact and pre-loads, this has important implications for the dependence of the small deformation stiffness that can be achieved on the structure's minimum feature size.  

Within the aforementioned bounds, we suggest that a wide range of qualitatively different nonlinear force-displacement paths, which can be fit by a polynomial (as a function of displacement/strain), should be achievable through this exploitation of geometry, even those which elsewhere in literature have been achieved through other mechanisms, mechanical or otherwise (\textit{e.g.}, magnetism, etc.). Further, while these qualitative design guidelines are helpful, we look herein to topology optimization techniques to refine the design process for more exact targeted nonlinear behaviors.

\subsection*{Optimization Method}

A 2D plane stress, geometrically nonlinear (large deformation), displacement controlled, level set optimization was implemented, following the method described in \cite{dunning_introducing_2015,chung_level-set_2020}. For simplicity, we restricted this work to a tension regime. A Kirchhoff material model was implemented, allowing for geometric nonlinearity and large strains while maintaining a linear elastic material model. We choose a level set optimization method because of its compatibility with our pursuit of simply manufacturable, binary (material and no material) geometries.

Within the context of the stiffness bounds described as part of Figure~\ref{Fig2}, we highlight that there exists an inherent coupling between the degree of nonlinearity and the stiffness of a structure (excepting phenomena related to snap through and snap back). We define degree of nonlinearity as the relative magnitude of nonlinear terms in the force-displacement relationship with respect to the linear term at a given displacement. To elucidate the coupling between nonlinearity and stiffness, we consider the transition between axial and bending deformation of a long thin beam. Taking a ratio of those two characteristic stiffnesses (axial by bending) for a beam of rectangular cross section, we obtain a ratio of $(Lt^{-1})^2$, demonstrating that thinner structures are capable of greater change in stiffness (\textit{i.e.}, nonlinearity) over a fixed displacement range. However, thinning a structure makes it softer, resulting in a uniformly lower magnitude force displacement curve. As mentioned above, many previous attempts at this problem have employed an objective function which attempts to directly target a force-displacement curve. This severely limits the design space, excluding multiple, and sometimes all, viable solutions. In that case, a better goal, in pursuit of a specific nonlinear behavior, is to search for geometries which will enable the desired strength and shape of nonlinearity, regardless of the stiffness of the structure. We address this by defining the objective function in terms of a polynomial that is fitted to the force-displacement curve, wherein the objective is set to match a desired ratio between polynomial coefficients. This objective is written as

\begin{equation}
    \text{min} \left( \frac{P_i}{P_j}\frac{T_j}{T_i} - 1 \right) ^2 + \left(P_k\right)^2,
    \label{Eq1}
\end{equation}

\noindent where $P_i$ and $P_j$ represent the current design polynomial coefficients being optimized for a desired ratio, $T_i$ and $T_j$ represent the target ratio, and $P_k$ is a current design polynomial being driven to zero. A target linear plus cubic polynomial constitutive law could thus be written as $F = T_jx + 0x^2 + T_ix^3$, while the current design's constitutive law would be written $F = P_jx + P_kx^2 + P_ix^3$, in which the ratio of the $i$ and $j$ terms would be optimized to match the target ratio, while the $k$ term would be driven to zero. For targets which have multiple desired ratios, multiple $j$ terms can be used. We note that, by optimizing for polynomial coefficient matching, a limitation is introduced, which restricts targeted nonlinearities to those which can be well-fit by a polynomial (this excludes behavior such as snap-back, for example, when using a polynomial that is a function of displacement). However, a wide variety of nonlinear curves can be well described by a polynomial fit, particularly as the order of the polynomial is increased, resulting in an extensive space of potential nonlinear targets to be explored.

Sensitivities for this objective function can be written as a combination of sensitivities for compliance and sensitivities related to the fitting of the polynomial. We calculate self-adjoint, nonlinear, displacement control compliance sensitivities, $\partial C_j/\partial \Omega$, as calculated in \cite{chung_level-set_2020}, where $C$ is the compliance and $\Omega$ is the domain. Calculating our polynomial fit through use of the equation $P = V^{-1}C$, where $V^{-1}$ is the inverse of the Vandermonde matrix, the sensitivity for each polynomial term is written as:

\begin{equation}
    \frac{\partial P_i}{\partial \Omega} = \sum_{j=1}^{m}\frac{\partial P_i}{\partial C_j}\frac{\partial C_j}{\partial \Omega} = \sum_{j=1}^{m} V^{-1}(i,j)\frac{\partial C_j}{\partial \Omega},
    \label{Eq2}
\end{equation}

\noindent where $i$ refers to the degree of polynomial, $j$ refers to the displacement step, and $m$ is the total number of displacement points under consideration. In addition, we employ a smoothing algorithm in the mapping of the sensitivities to the level set boundary nodes \cite{PICELLI20181}, which discourages very small, discontinuous notches and protrusions in the resulting design.

Despite the ability to identify gradients in the design space, we note that this particular inverse design problem is highly non-unique and non-convex. Hence, optimization can get stuck in local minima which may not be sufficiently close to the objective nonlinearity. In order to improve the quality of a solution, we use mechanical intuition to choose an initial condition that is qualitatively ``nearby'' the desired type of nonlinearity, through the heuristic spring design process described above. This narrows down the number of possible solutions the optimizer can reach, and reduces the chances of becoming trapped in an undesirable local minimum. A brief exploration of the effect initial condition choice can have on the ability to find a satisfactory local optimum is included in the SI. This approach, presented herein, requires some mechanical insight and designer input in choosing an initial condition. Similar approaches have been employed in other works, which have relied upon the input of reduced order training datasets \cite{deng_inverse_2022} and the use of multiple initial condition guesses \cite{li_digital_2022} to achieve the target constitutive behavior.

\subsection*{Examples}

To demonstrate the capability of the design method and optimization process, we optimize several spring structures to target specific nonlinear constitutive laws from several classes of nonlinear behavior, \textit{i.e.} softening, stiffening, softening to stiffening, and stiffening to softening. By optimizing a structure for each class of nonlinearity, we demonstrate the capability of the approach to span the nonlinear design space and target any type of nonlinearity that falls within the range of these classes. These results are depicted in Figure \ref{FigExamples}, in which the force-displacement equations with the following polynomial coefficients are targeted: Softening, $F = x - 40.0x^3$; stiffening, $F = x + 163.3x^3$; softening to stiffening, $F = 1.40x - 17.332x^2 + 166.875x^3$; and stiffening to softening, $F = 0.599978x + 17.332x^2 - 166.875x^3$. The displacement range was chosen arbitrarily, as were the coefficient weights, though in the latter two cases, the target polynomial coefficients were chosen to achieve symmetric stiffness switching springs, as outlined in \cite{chaunsali_self-induced_2019}. Nearly all targeted nonlinearities were matched to within less than 2\% of the desired coefficients (with the sole exception being the softening-to-stiffening case, discussed below). In the optimizations shown in Figure \ref{FigExamples}, the initial conditions were selected to be in the same class of nonlinearity as the desired target (for example, an initial condition which displayed a general stiffening nonlinearity was selected for the stiffening-type nonlinearities), with coarse adjustments being required to tune the initial polynomial coefficient signs. The initial conditions and polynomial fit details for each optimization case can be found in the SI. We suggest that all of the optimized designs can be considered ``non-intuitive'', in that they would be difficult or impossible to design by hand to the desired degree of precision using human intuition alone. We draw particular attention to the two stiffness switching optimized designs shown in Figure \ref{FigExamples} C and D, as these targeted constitutive laws were more complex in nature, requiring the optimizer to capture a qualitative shift in behavior and to perform optimization for two polynomial ratios simultaneously. The stiffening-to-softening spring design in Figure \ref{FigExamples} D, in particular, is of note in that it can be considered to be the combination of two springs (beams) in series (one stiffening spring and one softening spring) in order to achieve the desired behavior. These two springs are highly coupled, with the nonlinear behavior of each affecting not only the overall behavior of the structure, but also the boundary conditions and applied displacement of the other spring, creating a highly complex and nonlinear design scenario.  Although the fitness of the match is slightly lower for this case, at an error of 5.242\%, this is less surprising considering the greater structural complexity (and presumed design space complexity) seemingly required to achieve this response.

\begin{figure}[]
    \centering
    \includegraphics[width = \textwidth]{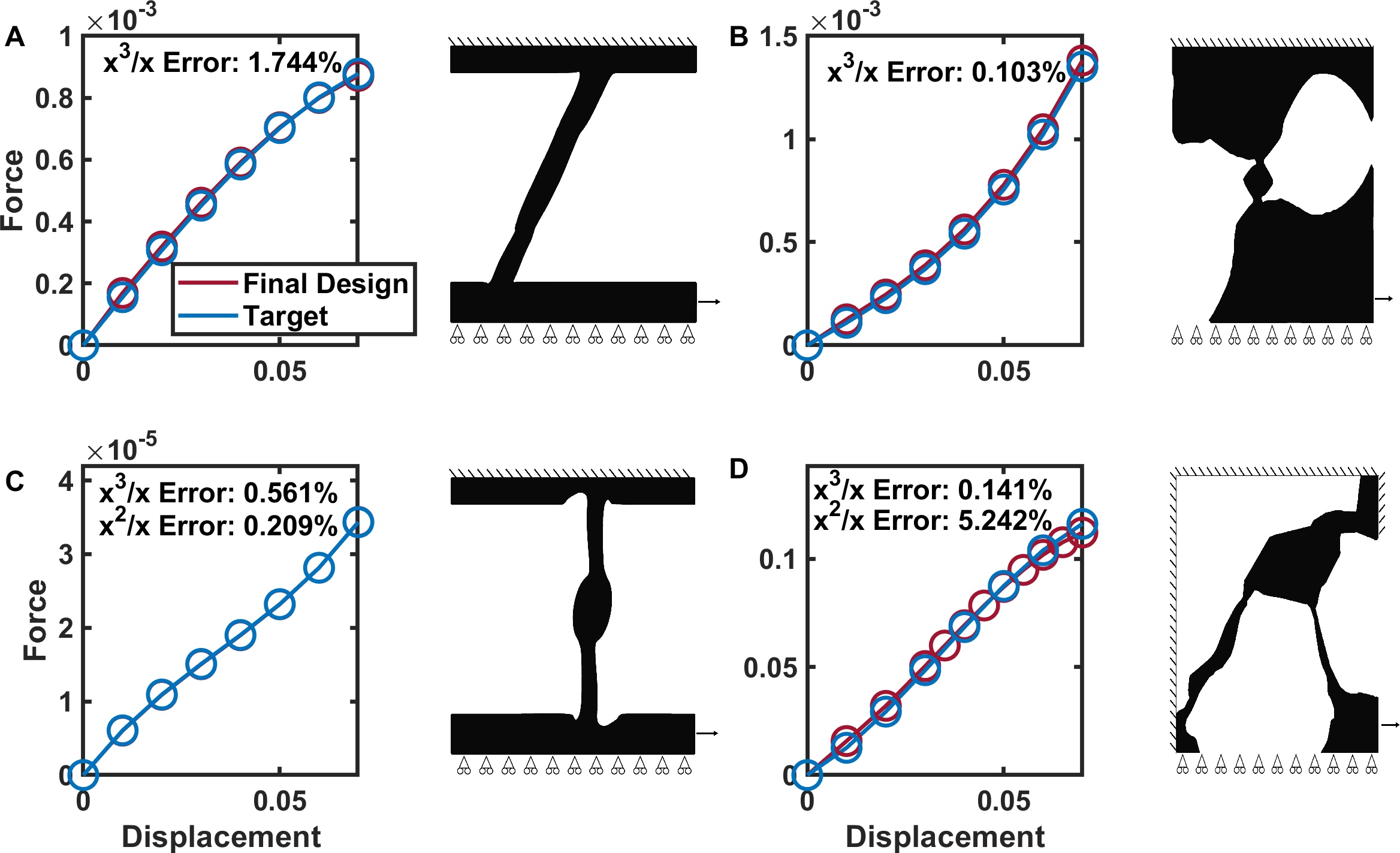}
   \caption{Optimized structures for four different classes of nonlinearity, including softening (A), stiffening (B), softening to stiffening (C), and stiffening to softening (D). The red curves represent the normalized polynomial law achieved by the design, while the blue represent the normalized target polynomial.}
    \label{FigExamples}
\end{figure}

The optimization results demonstrate that the method is capable of designing a variety of nonlinear springs. The high degree of precision in matching a target behavior also makes it possible to target nonlinearities which vary from one another only slightly, sweeping through the design space for the desired nonlinear behavior. To demonstrate this capability, we chose a target which consisted of a linear plus cubic stiffening type nonlinearity ($F = x + Ax^3$), and varied the target coefficient weights (A is varied). This type of sweep is fairly straightforward to perform once the first design has been optimized, as the optimum from one design can be used as the initial condition for the next step, in many cases (but not all) eliminating the need to design the initial conditions. The results are shown in Figure \ref{FigSweep}. We first draw attention to the distinct shift in design between ratios 40 and 80. This transition represents a local minimum which led to a relatively poor optimization of the ratio 40 target, and prevented an acceptable optimization of the ratio 20 design from an initial condition drawn from this poor ratio 40 optimization. Intuitively, this makes sense, as the hinge-like structure present in the more nonlinear designs is qualitatively different from the type of structure which might be expected to exhibit more linear behavior. Selecting a new starting design, which in this case consisted of straight edges and a thicker beam to promote a weaker nonlinearity, provided a successful workaround which avoided the local minimum and found a satisfactory design for the ratio 20 target, as well as an improved optimal design for the ratio 40 target. This difficulty highlights the challenges of the highly nonlinear design space, and the ease with which such local minima can be circumvented with educated initial design. We also note that the coupling of stiffness and nonlinearity is clearly visible in the optimized curves of the sweep designs (considering ratios 40-163 to be grouped together, and ratios 10-20 to be grouped together) in Figure \ref{FigSweep}B, in which the more linear designs display stiffer overall behavior, while the more nonlinear designs are softer. This change in nonlinearity (and stiffness) is achieved by increasing or decreasing the thickness. We highlight this in the zoomed-in images of the ratio 40-163 optimized structures in Figure \ref{FigSweep}, noting that the ratio 163 design is the optimized stiffening design from Figure \ref{FigExamples}, and that each subsequent ratio uses the optimum from the previous ratio as the initial condition.

With respect to the coupling between stiffness and nonlinearity, while, as in Figure \ref{MethodOverview}, it is simple to normalize or scale the linear stiffness (or resulting force magnitude) computationally, we also note that there are several mechanisms to achieve this in practice. Perhaps the most straightforward manner to scale the resulting stiffness or effective modulus would be to adjust the elastic modulus of the constituent material forming the design. However, in practice, materials occupy discrete moduli ranges and come with other practical limitations such as small elastic ranges. As an alternative approach, consider first the structural stiffness of the design. To keep the same degree of nonlinearity, while increasing the magnitude of the force (thus compensating for the normalized stiffness), one could change the depth of the 2D geometry, as shown by the varied colors in Figure \ref{MethodOverview}C and the resulting force-displacement curves in Figure \ref{MethodOverview}D. One could also scale the area of the 2D geometry, thus increasing the force, however, the applied displacement must also be proportionally scaled to achieve the same degree of nonlinearity. We next consider scaling the response from the perspective of an effective stress-strain response. Consider the 2D design with a nonlinear stiffness $k(\delta)$ is assigned some depth $d$, length $l$, and width $w$, as shown in Figure \ref{MethodOverview}C. This structure is then repeated within a larger unit-cell domain of depth $D$, length $L$, and width $W$, wherein the designed region is attached to a rigid frame as shown in Figure \ref{MethodOverview}E. The resulting stress strain relationship would thus be $\sigma =\epsilon kL(DW)^{-1}$, such that the size of the unit cell becomes essentially a single scaling factor equivalent to scaling the modulus of the constituent material. There is a key difference from scaling the constituent modulus, however, where the lengths defining the design domain cannot exceed that of the larger, unit cell domain.

\begin{figure}[]
    \centering
    \includegraphics[width = 0.5\textwidth]{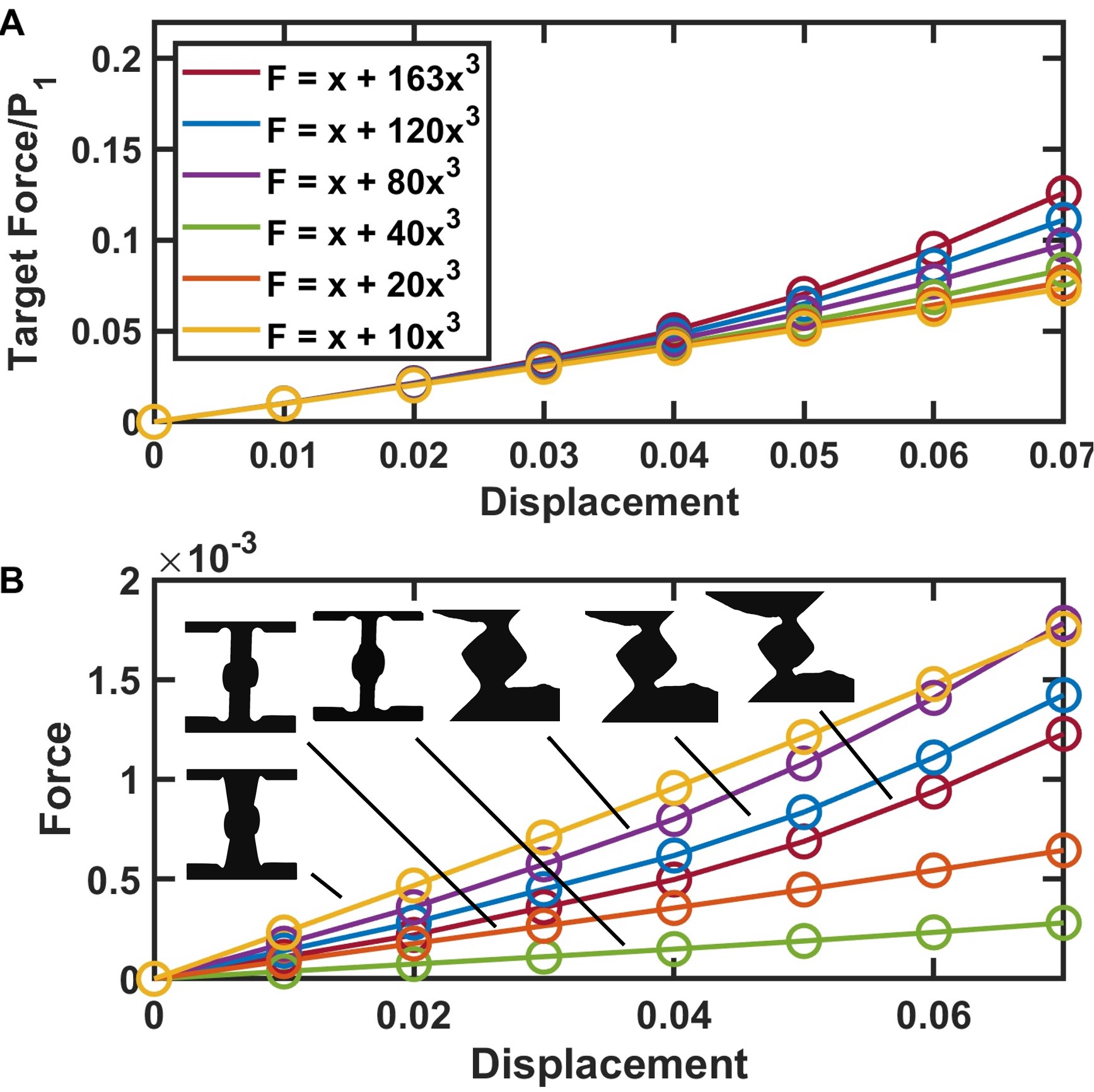}
   \caption{Optimized springs resulting from a sweep of the nonlinear term in the force displacement equation $F = x + ax^3$, in which $a$ is varied from 10 to 163. A) Normalized results, wherein the linear stiffness has been divided out such that all curves have a uniform linear stiffness. B) Non-normalized results, in which curves have been left with the varied linear stiffness directly from the optimization results, highlighting the stiffness variation that is exploited in the optimization process to achieve varying degrees of nonlinearity. The non-normalized force-displacement equations are as follows, in order of descending nonlinearity: $F = 0.0107 + 1.753x^3$, $F = 0.0118x + 1.4041x^3$, $F = 0.0156x + 1.2417x^3$, $F = 0.00377x + 0.1498x^3$, $F = 0.0089x + 0.1784x^3$, $F = 0.0232x+0.232x^3$.The snapshots of the three optimized structures on the right (target ratios of 80 to 163) are zoomed in on the high sensitivity regions. The full structures can be found in the SI.}
    \label{FigSweep}
\end{figure}

Furthermore, we include several additional examples of optimizations in order to demonstrate the range and versatility of the method. While the examples above highlight distinct nonlinear behaviors, several of the targeted curves are weakly nonlinear. To demonstrate that the method is not constrained to weak nonlinearities, Figure \ref{FigNewOpts}A highlights a targeted stiffening curve displaying a greater degree of nonlinearity (and a purely quadratic constitutive law). The target in this case is $F = x + 200.0x^2 + 0x^3$. We note that, when the maximum displacement of 0.07 is substituted into $x$ in this target, the ratio of the quadratic term to the linear term becomes 14, demonstrating an order of magnitude difference between linear and nonlinear terms. In Figure \ref{FigNewOpts}C, a constitutive law containing a fourth order term is targeted in order to show that the above cubic examples are an arbitrary choice of nonlinearity, and there is nothing constraining the method to third order terms. In addition to targeting the distinct nonlinear behaviors outlined above, we desire the capability of targeting nonlinearities which lie in the negative stiffness, as well as fully bistable regime (a capability which is highly useful in pursuing superior impact-resistant structures.) \cite{shan_multistable_2015} One benefit of our displacement-controlled approach is the capability to solve problems which exhibit negative stiffness and snap-through type behavior with a Newton-Raphson solver. In Figure 5B below we show optimizations which target a fully bistable forced-displacement curve. Results of an optimized negative stiffness (but not fully bistable) structure are shown in Figure 5D, for a targeted nonlinear law of $F = x - 10.0x^2 + 30.004x^3$. For both of these cases, we optimize over a greater displacement range in order to more easily realize the more nonlinear negative stiffness behavior without resorting to a significantly increased mesh density (as would be required for a thinner structure). As above, the displacement range and targeted nonlinearity are otherwise arbitrary. This optimization requires the simultaneous optimization of two ratios. For the initial optimization run, both ratios were matched to within 1\% (0.638\% for the cubic and 0.636\% for the quadratic). However, in this case, the negative stiffness portion of the curve is heavily dependent upon only one of these two ratios (the quadratic, which is responsible for the transition to negative stiffness due to the negative sign associated with the term), and the behavior in the negative stiffness region is accordingly much more sensitive to this ratio. We therefore add a weight (a 10x multiplier) to the quadratic ratio for this optimization, prioritizing this term over the cubic term. This result (shown in Figure \ref{FigNewOpts}D) achieved a lower match to the cubic term (1.712\%), but a much closer match to the prioritized quadratic term (0.180\%), resulting in a closer normalized match of the negative stiffness portion of the curve. 

\begin{figure}[]
    \centering
    \includegraphics[width = \textwidth]{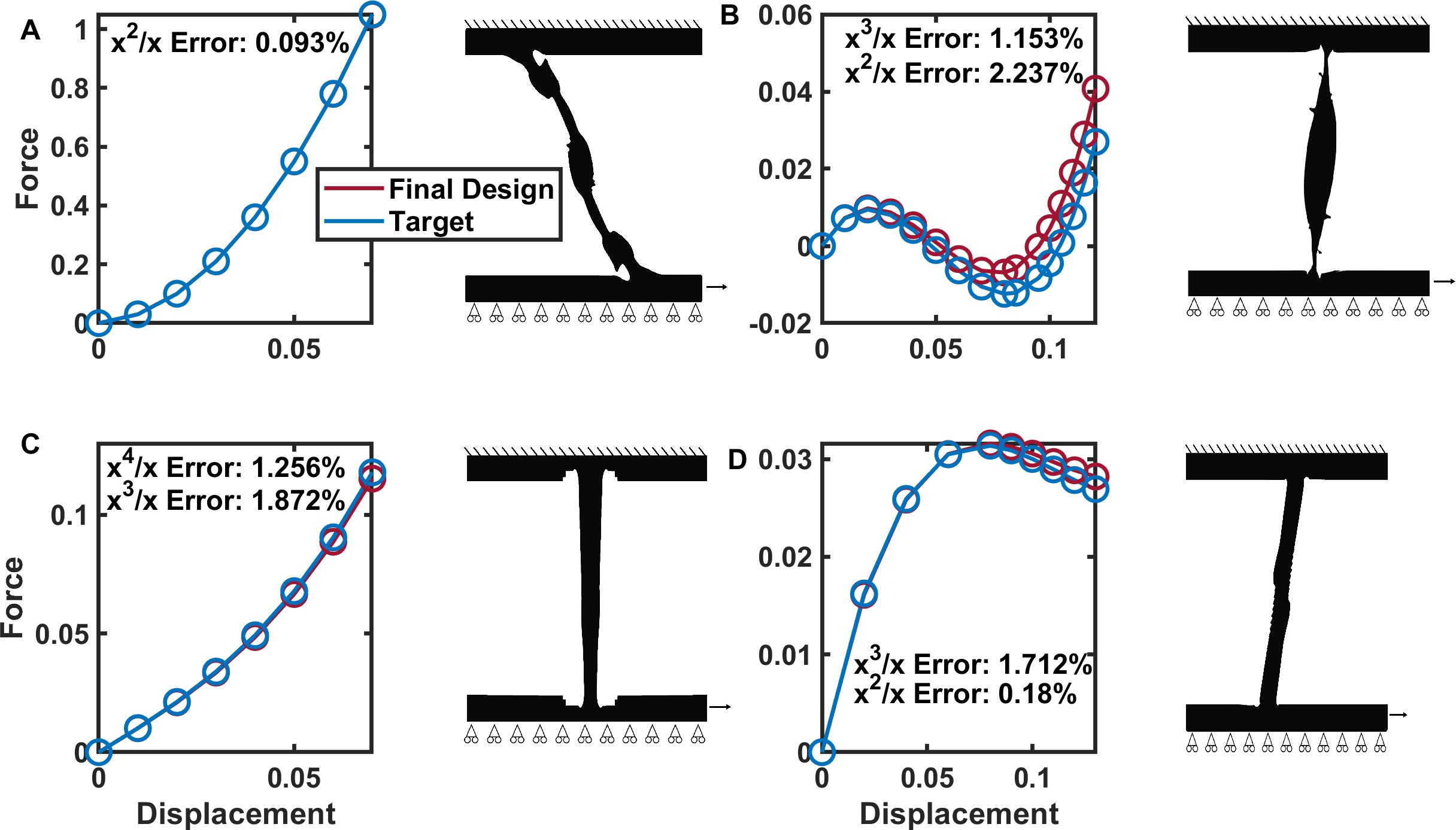}
   \caption{Additional optimization examples, highlighting the versatility of target-able curves. A) A purely quadratic, highly nonlinear stiffening optimization. B) A bistable optimziation. C) A fourth order optimzation. D) A negative stiffness (but not fully bistable) optimziation.}
    \label{FigNewOpts}
\end{figure}

\subsection*{Experimental Verification}

In order to confirm the accuracy of the method and the validity of the optimized springs, several of the springs shown herein were chosen for manufacturing and experimental verification of the nonlinear force-displacement curves. Polycarbonate was initially chosen to fabricate the springs from, due to its relatively linear behavior at lower strains and high strain-to-break value. However, preliminary tests indicated that for thicker structures, plasticity onset (which is not accounted for in the optimization's Kirchhoff material model) would introduce unaccounted nonlinear behaviors. The thinnest structure, the bistable spring from Figure \ref{FigNewOpts}B, was therefore manufactured and tested in polycarbonate, with the results shown in Figure \ref{FigExp}A indicating close agreement between optimization outputs and experiments.  It is of note that such experimental validation proved to be incredibly sensitive to misalignment issues (such as any out of plane offset or the direction of displacement application not being precisely horizontal) and quality of the boundary condition approximation. Regarding the latter, the imposition of the y-direction being fixed on the top and bottom edges via the frame containing the optimized structure, in particular, was found to be very important, due to a great deal of the nonlinearity stemming from the structure pressing against these surfaces during deformation. Indeed, insufficient stiffness of the frame, or lateral forces applied to it, were observed to modify the resulting nonlinear force-displacement response of the designed structure. Careful construction of the system was thus required in order to achieve good experimental match, which may require additional attention when considering real-world applications. We also note that, due to the FE simulations being run on a 1x1 m domain, often with a modulus of E = 1 Pa, simulation results required scaling to match up to experimental results. This is to be expected, and highlights the ability to manipulate the scale of the structure in order to achieve desired force ranges. Simulation results were scaled along the x-axis by the size of the unit cell (\textit{i.e.}, experimental unit cell length/simulation unit cell length), and along the y-axis by the modulus of the constituent material used in the experiment (found via scaling max and min values of nonlinear curves, as described in the SI), the edge length of the unit cell, the depth, and a factor of two to account for two springs being present in the experiment(\textit{i.e.} experimental unit cell length/simulation unit cell length), and along the y-axis by the modulus of the constituent material used in the experiment (found via scaling max and min values of nonlinear curves, as described in the SI), the edge length of the unit cell, the depth, and a factor of two to account for two springs being present in the experiment (\textit{i.e.} ($E_{Exp}/E_{Sim})*(L_{Exp}/L_{Sim})*(D_{Exp}/D_{Sim})*2$, in which \textit{E} refers to Young’s Modulus, \textit{L} refers to the edge length of the unit cell, \textit{D} refers to the depth of the unit cell, subscript Exp refers to the value in the experiment, and subscript Sim refers to the value used in the optimization).

As the ability to test any optimized structure was desired, an updated material model was introduced to the modeling for optimization, which would allow for the use of fabrication materials that exhibit a larger elastic range. A Neo-Hookean (hyperelastic) material model was introduced, with a plane strain formulation. Several of the above optimization cases were re-run with the updated material model, beginning from the optimal configuration identified with the Kirchhoff material model. Small adjustments were made to these designs by the optimizer to account for the new material model. These updated structures were then fabricated out of silicone (see SI for material and fabrication details) and tested. The results, shown in Figure \ref{FigExp}B,C, demonstrate that the updated material model has similar levels of agreement to the original Kirchhoff material model, and allows for the fabrication and testing of thicker structures while avoiding the onset of plasticity. We note that analogous scaling was performed to map between simulation and experimental results, as described above in the elastic case. Further information on experimental validation is included in the SI.

\begin{figure}[]
    \centering
    \includegraphics[width = \textwidth]{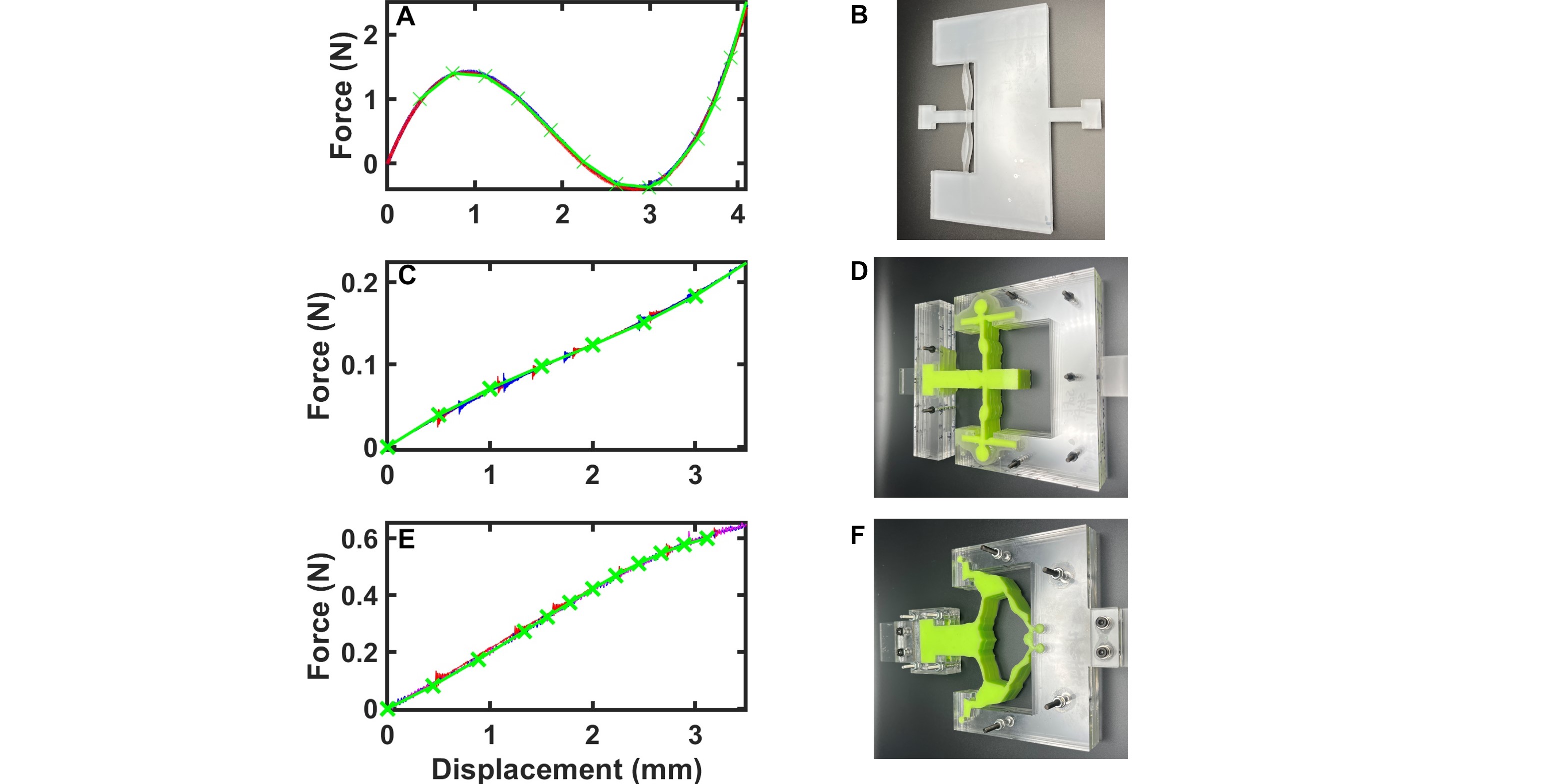}
   \caption{EExperimental validation of optimized spring force-displacement curves, in which red and blue solid lines indicate experimental data, and green lines with x markers are overlaid simulated force-displacement curves. Shown are A-B) the bistable optimized spring shown in Figure \ref{FigNewOpts}B (fabricated from polycarbonate, in which the frame has dimensions from top edge to bottom edge of 134 mm), C-D) the softening-to-stiffening spring shown in Figure \ref{FigExamples}C (fabricated from silicone, in which the frame has dimensions from top edge to bottom edge of 140 mm), and D-E) the stiffening-to-softening spring shown in Figure \ref{FigExamples}D (fabricated from silicone, in which the frame has dimensions from top edge to bottom edge of 120 mm).}
    \label{FigExp}
\end{figure}

\section{Conclusion}

The optimization design methodology developed herein targets polynomial material or structural constitutive response coefficient ratios, decoupling the desired nonlinear behavior from the inherent stiffness in its description. Through this, we bypass the limitations associated with stiffness-nonlinearity coupling. We have demonstrated the capability to design nonlinear springs with both breadth and precision. With desired high-precision nonlinear behavior having importance in areas ranging from impact mitigation to wave tailoring metamaterials, compliant mechanisms, mechanical logic and signal processing, soft robotics, and biomedicine (particularly settings involving biological-synthetic interfaces), we envision this approach enabling the development of new designer materials within these areas. 

Future near-term potential applications of this method may include the experimental implementation of several 1D chains, as proposed in Figure \ref{MethodOverview}E (\textit{e.g.} \cite{chaunsali_self-induced_2019, PhysRevE.95.062216} and impact-mitigating mechanical metamaterials. While 1D chains are an area of rich physics in which to validate the implementation of the method described herein, the extension to 2D, and furthermore three-dimensional (3D), are expected to present a host of new challenges. A system in which each unit cell interacts nonlinearly with the one next to it, and in which the nonlinear response is expected to be characterized by multiple degrees of freedom, introduces an increased degree of complexity which bears future examination. Such advancements will help enable tailoring of 2D or 3D bulk materials for loading in multiple directions. 

\medskip
\textbf{Data Availability Statement} \par 
Data supporting the results presented herein can be found in Mendeley Data, V1, doi: 10.17632/yp6cychpgr.1. This data includes .dxf files of the final optimized designs, as well as text files containing the raw output (force, displacement, and time) from the experiments. Further data can be provided from the authors upon reasonable request.

\medskip
\textbf{Acknowledgements} \par
N.B. acknowledges useful discussions with F. Talke, and both F. Talke and K. Loh for sharing their mechanical testing equipment. B.M., K.Q., and N.B. acknowledge support from the US Army Research Office (Grant No. W911NF-20-2-0182). H.A.K. and N.B. acknowledge support from the UC National Laboratory Fees Research Program of the University of California, Grant Number L22CR4520. B.M. acknowledges support from the U.S. Department of Energy (DOE) National Nuclear Security Administration (NNSA) Laboratory Graduate Residency Fellowship (LRGF) under Cooperative Agreement DE-NA0003960. I.F. acknowledges support from the Department of Defense (DoD) through the National Defense Science \& Engineering Graduate (NDSEG) Fellowship Program. A.P. and A.S. acknowledges support from the ENLACE summer research program at UC San Diego.

\bibliographystyle{ieeetr}
\bibliography{bibliography.bib}

\end{document}